\documentclass[twocolumn,prb,superscriptaddress,aps]{revtex4-2}
\usepackage[english]{babel}
\usepackage[utf8]{inputenc}
\usepackage{lmodern}

\usepackage{dcolumn}
\usepackage{amssymb, amsmath}
\usepackage{graphicx}
\usepackage{graphics}
\graphicspath{{figures/}} 

\usepackage{bm}
\usepackage{xcolor}
\usepackage{hyperref}
\usepackage{microtype}
\usepackage{braket}
\usepackage{physics}

\newcommand{\beq}{\begin{eqnarray}}
\newcommand{\eeq}{\end{eqnarray}}

\begin{document}

\title{Magnetic anisotropy from interligand hopping in strongly correlated insulators: application to the magnon spectrum of CrI$_3$}

\author{Evgenii Barts}
\email{e.barts@rug.nl}
\affiliation{RIKEN Center for Emergent Matter Science (CEMS), Wako, Saitama 351-0198, Japan}
\affiliation{Zernike Institute for Advanced Materials, University of Groningen, Nijenborgh 3, 9747 AG Groningen, The Netherlands}

\author{Paolo Barone}
\affiliation{CNR-SPIN Institute for Superconducting and other Innovative Materials and Devices, Area della Ricerca di Tor Vergata, Via del Fosso del Cavaliere 100, I-00133 Rome, Italy}

\author{Maxim Mostovoy}
\affiliation{Zernike Institute for Advanced Materials, University of Groningen, Nijenborgh 3, 9747 AG Groningen, The Netherlands}
\date{\today}

\begin{abstract}

Spin-orbit coupling (SOC) gives rise to complex magnetic states such as spin liquids, skyrmion crystals, and topological spin-wave excitations.
We consider exchange interactions in multi-orbital Mott insulators where SOC is strong on {\em ligand} ions.
SOC on the ligands enables electron hopping accompanied by spin flips and fluctuations in the orbital state of the ligand hole. These processes generate anisotropic exchange interactions and greatly increase the number of possible exchange paths.
The number grows further with the inclusion of hopping between ligands, which mediates interactions between more distant spins.
We propose an effective method to calculate exchange interactions at arbitrary separations between spins. Applying it to monolayer CrI$_3$, we obtain anisotropic interactions between nearest-neighbor and next-nearest-neighbor Cr spins, as well as single-ion anisotropy induced by long-range hopping.
In this material, magnetic anisotropy stabilizes long-range ferromagnetic order and opens a magnon gap at the Dirac points, which defines a nontrivial magnon band topology.
Using Hubbard model parameters from first-principles calculations, the resulting spectrum agrees well with the spin-wave dispersion observed experimentally in bulk CrI$_3$, except that the calculated Dirac gap is much smaller.

\end{abstract}

\maketitle

\section{Introduction}

\begin{figure*}[htb]
    \centering
\includegraphics[width=0.99\linewidth]{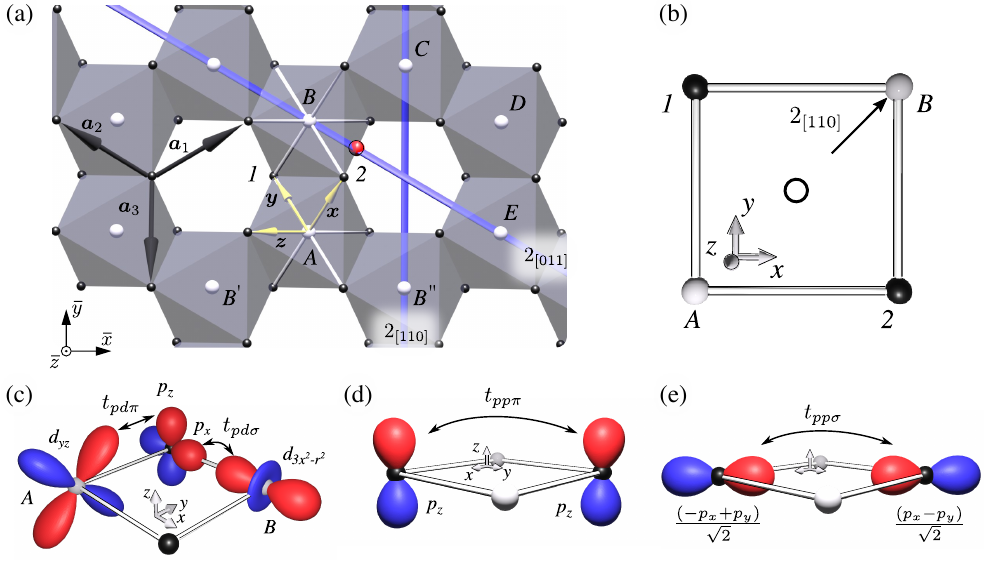}
\caption{\label{ch5:fig:CrI3lattice} { CrI$_{3}$ monolayer crystal and electronic structure.} 
    (a)
    The white (black) spheres show Cr$^{3+}$ (I$^{-}$) ions.
    Blue lines indicate the two-fold rotational symmetry axes of the CrI$_3$ monolayer, $2_{[110]}$ and $2_{[011]}$ (we use the Cartesian ${ x}, { y}$ and ${ z}$ axes shown by yellow arrows).
    The red point indicates the inversion symmetry center of an ideal iodine lattice, important for exchange interactions between the Cr ions $A$ and $C$.
    (b)
    Symmetries of an ideal Cr-I plaquette in the $xy$-plane: two-fold rotation axes, $2_{[110]}$ and $2_{[001]}$, and inversion around the center of the plaquette.
    {(c)} $t_{pd\pi}$ is the amplitude of hopping between the $d_{yz}$ orbital of the metal ion $A$ and the $p_z$ orbital of the ligand ion 1; $t_{pd\sigma}$ describes the hopping between the $p_x$ and $d_{3x^2 - r^2}$ orbitals.
    {(d)},~(e) The amplitudes $t_{pp\pi}$ and $t_{pp\sigma}$ describe the hopping between $p$-orbitals of neighboring ligand ions.
    }
\end{figure*}

\begin{figure*}[htb]
    \centering
\includegraphics[width=0.99\linewidth]{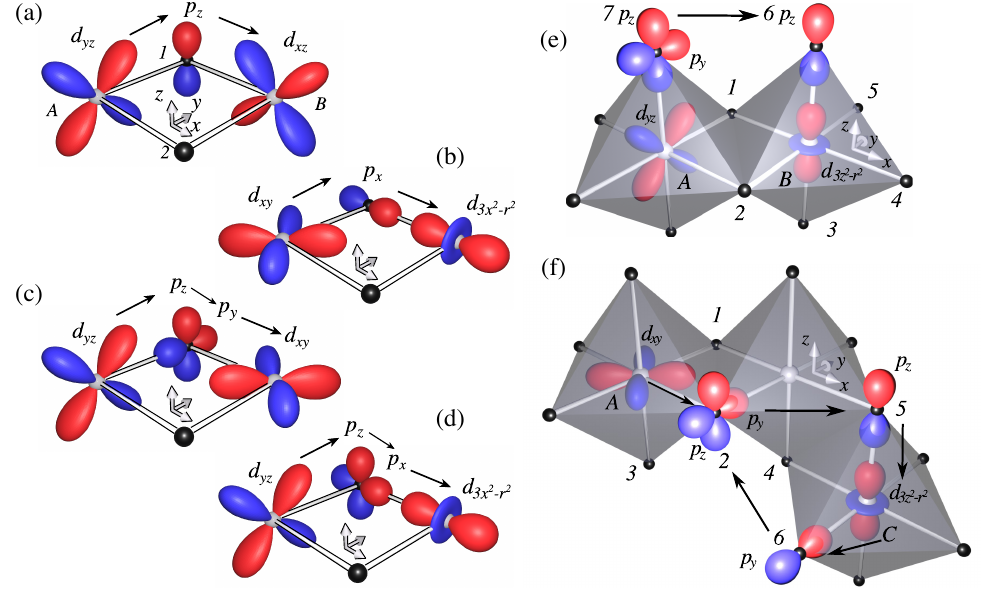}
    \caption{\label{fig:hoppings} {Effective $dd$-hopping}. 
     Isotropic Heisenberg exchange interactions from electron hopping between {(a)} $d_{yz}$ and $d_{xz}$ orbitals via $p_z$ ligand orbital and
    {(b)} $d_{xy}$ and $d_{3x^2 -r^2}$ orbitals via $p_x$ orbital. 
    White (black) spheres show Cr$^{3+}$ (I$^{-}$) ions and the hopping direction is indicated by arrows.
     An electron hops with spin flip {(c)} between the $d_{yz}$ and $d_{xy}$ orbitals via the SOC-entangled $p_z$ and $p_y$ ligand orbitals, and
    {(d)} between the $d_{yz}$ and $d_{3x^2 -r^2}$ orbitals via the SOC-entangled $p_z$ and $p_x$ ligand orbitals. 
    {(e)} 
    An electron hops from the $p_z$ orbital of ligand 6 to the empty $d_{3z^2-r^2}$ orbital of Cr-$B$; the ligand hole at site 6 is filled by the electron from the ligand 7 $p_z$ orbital; finally, the electron occupied the $d_{yz}$ orbital of Cr-$A$ hops into the $p_z$ orbital of ligand 7 via the  $p_y$ orbital with the spin flip due to the SOC.
    {(f)} 
    Hopping from the $d_{xy}$ orbital on site $A$ to the $d_{3z^2-r^2}$ orbital on site $C$ due to the tunneling between the $p_z$ orbitals of ligand ions 2 and 5. It is preceded by a spin flip on site 2, followed by a change of the orbital state from $p_y$ to $p_z$. The electron can hop back to site $A$ via the $p_y$ orbitals of ligand ions 2 and 6.
    }
\end{figure*}

\noindent
Building on the breakthroughs achieved with graphene, two-dimensional van der Waals magnetic materials have emerged as a platform for the electrical and optical control of magnetic order~\cite{Geim2013,Gong2017, Burch2018, Gibertini2019,Soriano2020, McGuire2017}.
In transition metal trihalides such as FeCl$_3$ and CrX$_3$ (X=Cl, Br, I), 
octahedrally coordinated transition metal ions form $ab$ layers with a honeycomb spin lattice [see Fig.~\ref{ch5:fig:CrI3lattice}(a)]. 
Among these compounds, CrI$_3$ stands out for its strong out-of-plane magnetic anisotropy, primarily related to the spin-orbit coupling (SOC) of iodine ions~\cite{Lado2017, Xu2018,Kvashnin2020, Jaeschke-Ubiergo2021}.
As in RuCl$_3$ with strong SOC on magnetic Ru sites, materials with strong SOC on ligand sites are promising candidates for realizing the fascinating physics of the Kitaev model, which originates from strong bond-dependent anisotropic exchange interactions~\cite{Kitaev2006, Jackeli2009, Plumb2014, Kim2015}.

Uniaxial magnetic anisotropy opens a gap in the spin-wave spectrum at the $\Gamma$ point and stabilizes the ferromagnetic order observed in a single layer of CrI$_3$~\cite{Huang2017}, but it is insufficient to describe magnetism in this material.  
The spin-wave spectrum also shows a large gap (of a few meV) at the Dirac points~\cite{Chen2018}, indicating nontrivial topology of magnon bands and the presence of protected magnon edge states~\cite{Mook2021, Wang2021}. 
On the basis of angular-dependent ferromagnetic resonance measurements, this gap was ascribed to Kitaev interactions that are much  stronger than Heisenberg interactions between nearest-neighbor spins~\cite{Lee2020}.   
Alternatively, a model with Dzyaloshinskii-Moriya interactions (DMI) between next-nearest-neighbor spins gives a good description of the spin-wave spectrum measured under in-plane magnetic fields~\cite{Chen2021a}.
However, the anisotropic exchange interactions obtained from {\em ab initio} calculations are too weak to explain the observed gap at the Dirac points~\cite{Lado2017, Xu2018,Kvashnin2020, Jaeschke-Ubiergo2021}.
Perturbative calculations for the extended Hubbard model, which includes both $3d$ electrons on transition metal sites and $5p$ electrons on ligand sites, give  
Kitaev interactions that are weak compared to ferromagnetic Heisenberg interactions between nearest-neighbor spins~\cite{Stavropoulos2021}.

The case of CrI$_3$ illustrates the need for 
accurate calculations of exchange interactions in Mott-Hubbard insulators, especially in those with strong SOC on ligand sites.
Electron hopping between two transition metal ions can change the spin projection on the ligand ion, leading to both symmetric and antisymmetric anisotropic exchange interactions~\cite{Moriya1960}.  
Such spin flips are accompanied by a change in the orbital state of the ligand hole, which allows for exchange processes not accounted for by the Goodenough-Kanamori (GK) rules~\cite{Goodenough1955,Kanamori1959}.
For example, virtual processes involving hopping between $t_{2g}$ orbitals of neighboring Cr ions produce antiferromagnetic Heisenberg exchange interactions, and hopping between $t_{2g}$ and $e_{g}$ orbitals contributes to ferromagnetic exchange interactions [see Figs.~\ref{fig:hoppings}(a) and (b)]. This is in agreement with the GK rules.
By contrast, spin-flip hopping combined with a change of the orbital state of the ligand hole [see Figs.~\ref{fig:hoppings}(c) and (d)] leads to Kitaev and other anisotropic exchange interactions beyond the GK rules.

In this paper, we extend the perturbative calculations of Ref.~\onlinecite{Stavropoulos2021} by including hopping between ligand $p$ orbitals in the $dp$ model.  
First, interligand hopping enables the calculation of exchange interactions between distant spins.
Second, it lowers the local symmetry of the transition metal sites even when symmetry-allowed distortions of the ligand octahedra are neglected.
We show that the effect of the $pp$-hopping on anisotropic exchange interactions is comparable to that of lattice distortions reported in Ref.~\onlinecite{Stavropoulos2021}.
We calculate the anisotropic spin-spin interactions for the monolayer CrI$_3$ while neglecting distortions of the octahedra formed by I ions.

Exchange interactions between distant spins involve many exchange paths, which complicates their calculation.
We propose the following procedure that automatically accounts for all possible exchange paths and works for arbitrarily large separations between spins: 
(i) compute the band structure of $p$ electrons that hop in the sublattice formed by the ligand ions;
(ii) calculate the hopping amplitudes between ligand and transition metal ions in an intermediate state with one more or one less electron;
(iii) for a pair of transition metal sites, derive effective hopping amplitudes $t_{dd}$ which depend on the intermediate states of both ions;  
(iv) obtain the spin Hamiltonian in the effective $d$-electron model to second order in $t_{dd}$.

We perform \textit{ab initio} calculations to obtain the $dp$ and $pp$ hopping amplitudes, while the other parameters of our microscopic model are taken from the experimental and theoretical literature.
The resulting spin Hamiltonian is used to compute the magnon dispersion.
The bandwidths of optical and acoustic magnons and the energy gap at the $\Gamma$ point are in good agreement with experimental data on bulk CrI$_3$.
This gap results primarily from uniaxial anisotropy, which can be understood in terms of an effective SOC on transition metal sites induced by the admixture of states containing ligand holes.  
We also find that the DMI between next-nearest-neighbor spins is zero because the undistorted network of ligand ions preserves the inversion symmetry, and that anisotropic exchange interactions between both nearest- and next-nearest-neighbor spins are too weak to explain the experimentally observed gap at the Dirac points.

\section{Microscopic Hubbard model}
\label{sec:micro}

\begin{figure}[htb]
    \centering
\includegraphics[width=0.99\linewidth]{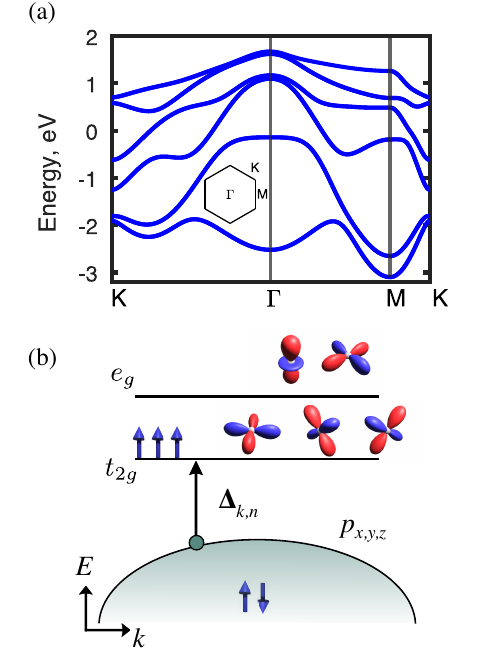}
    \caption{\label{ch5:fig:CrI3bands}
    {(a)}~The iodine sublattice energy band structure, shown for high-symmetry lines.
    Zero energy corresponds to the energy of isolated  iodine's $p$ orbital state. 
    Hopping amplitudes are $t_{pp\pi}=0.15$~eV and $t_{pp\sigma}=0.7$~eV; SOC is $\lambda = 0.63$~eV.
    {(b)}~Illustration of electron virtual hopping from fully-occupied $p$ orbital I bands to localized $d$ Cr states.
    }
\end{figure}
\noindent
For our microscopic calculations of CrI$_3$ spin model parameters, we use the $dp$ model which describes the $d$ orbitals of transition metal ions and the $p$ orbitals of ligand ions~\cite{Feldkemper1998}. The zeroth-order model Hamiltonian describes isolated Cr and I ions, as well as the hopping between the ligand sites,
\begin{equation}
\label{eq:Hubbard}
    H_{0} = H_{\rm Cr} + H_{\rm I} + H_{\rm pp}.
\end{equation}
Here, $H_{\rm Cr}$ describes electrons occupying the $t_{2g}$ and $e_g$ orbitals of Cr ions with energies $\varepsilon_{t_{2g}}$ and $\varepsilon_{e_g}$, respectively.  In the unperturbed $3d^3$ electron configuration, the $t_{2g}$ shell is half-filled and the $e_g$ orbitals, with higher energy due to the crystal field splitting, $\Delta_c = \varepsilon_{e_{g}}-\varepsilon_{t_{2g}}$, are empty. The on-site Coulomb interactions between electrons are described by the Kanamori Hamiltonian~\cite{Kanamori1963}. The relatively weak SOC on Cr sites is neglected.

The term $H_{\rm I}$ describes $5p$ orbitals of iodine ions, which are split by SOC of strength $\lambda$ into a doublet and a quadruplet with energies $\varepsilon_{p} - \lambda$ and $\varepsilon_{p} +\lambda/2$, respectively. Since we only consider intermediate states with a single ligand hole, the Coulomb repulsion on ligand sites can be absorbed into the definition of electron energy, $\varepsilon_{p}$. We ignore distortions of iodine octahedra, but instead take into account the hopping between nearest-neighbor ligand sites. The delocalization of electrons over the network of I ions results in electron band states, $\ket{V_{\bm k,n}}$, where $\bm k$ and $n$ are the wave vector and the band index. 

The perturbation is electron hopping between the $d$ and $p$ sites, described by the Hamiltonian 
\begin{equation}
\label{eq:TB}
\begin{split}
     H_{\rm dp}=  \sum_{\langle i\alpha,j\beta\rangle,\sigma}
    \biggl[&t_{i\alpha,j\beta}^{pd} \left(d_{i\alpha\sigma}^\dagger p_{j\beta\sigma} + \mathrm{h.c.}\right) \\
    +\, &t_{i\alpha,j\beta}^{pp} \left(p_{i\alpha\sigma}^\dagger p_{j\beta\sigma} + \mathrm{h.c.}\right) \biggr]
 \, ,
    \end{split}
\end{equation}
where $p_{i\alpha\sigma}$ annihilates an electron with spin projection $\sigma$ in orbital $\alpha$ of the I ion at site $i$. It might seem strange that $H_{\rm pp}$, describing the interligand hopping with a relatively small hopping amplitude, is included into $H_0$, whereas $ H_{\rm dp}$ is considered as a perturbation. This is done in order to calculate exchange interactions between further-neighbor spins. The expansion of our results in the ratio of the $pp$-hopping amplitudes and charge-transfer energy gives the conventional perturbation expansion in all hopping amplitudes. However, the number of exchange paths grows rather quickly with the distance between spins [examples of exchange paths resulting from the $pp$-hopping are shown in Figs.\ref{fig:hoppings}{(e)} and {(f)}]. The $p$-bands provide an efficient way to calculate the sum over all paths. 

The hopping processes, shown in Fig.~\ref{ch5:fig:CrI3lattice},  are parametrized using the Slater-Koster scheme~\cite{Slater1954}. The amplitudes $t_{pd\pi}$ and $t_{pd\sigma}$ describe the hopping between neighboring transition metal and ligand sites (panel~{(c)}), and $t_{pp\pi}$ and $t_{pp\sigma}$ describe the hopping between neighboring ligand sites (panels~{(d)} and {(e)}).
In order to get realistic estimates of hopping amplitudes, we resort to {\em ab initio} Density Functional Theory (DFT) calculations. By projecting the band structure onto maximally localized Wannier functions for Cr and I atoms, we obtain full $pd$-, $pp$- and $dd$-hopping Hamiltonians between $p$ and $d$ orbitals (see Appendix~\ref{app:dft} for details). 
We neglect smaller direct $dd$ hopping amplitudes, whereas our calculations reveal remarkably large values of $pp$ hopping amplitudes, which have not been addressed in the literature before.
The Slater-Koster parameters, estimated from the hopping integrals of the Wannier tight-binding model between Cr and I atoms within a plaquette (see Fig.~\ref{ch5:fig:CrI3lattice}{(b)}), are $t_{pd\sigma}=1~$eV, $t_{pd\pi}=0.5~$eV, $t_{pp\sigma}=0.7~$eV and $t_{pp\pi}=0.15~$eV.

We adopt the remaining model parameters from previous research.
The Hubbard constants are $U=3$~eV and $U'= U - 2J_{\rm H}$, a commonly used approximation. The crystal field splitting energy is $\Delta_c = 1$~eV, and the atomic SOC on iodine sites is $\lambda=0.63$~eV. 
$\Delta_c$ corresponds to the energy of the $dd$ electron transition between Cr $t_{2g}$ and $e_g$ orbitals, which can be attributed to the experimentally observed photoluminescence peak at $1.1$~eV~\cite{Seyler2018, Jin2020}.
These $dp$ model parameters are consistent with those used in earlier {\it ab initio} studies~\cite{Kvashnin2020,Jang2019,Soriano2021,Yadav2024}.
Our Hund's rule coupling is $J_{\rm H}/U =0.1 $, which, although lower than 0.24 suggested by constrained random phase approximation (cRPA) calculations~\cite{Jang2019}, agrees with recent  cRPA calculations~(0.11) that include environmental screening effects~\cite{Soriano2021}.
Our calculations show that the charge-transfer energy is given by $\Delta_{\rm CT}  = 1.5$~eV for these parameters. This is the minimum energy required to transfer an electron from an I site to the Cr site: ${\rm min}\left(\Delta_{S=2}^{e_{g}}\right)_{\bm k}$, which can be associated with the experimentally observed light absorption peak close to 2~eV~\cite{Seyler2018, Jin2020}. The charge transfer energy is the last parameter that can be treated as a free parameter in our microscopic model, as the energy of a single hole on a ligand can vary significantly from its bare value, such as $\varepsilon_{t_{2g}}-\varepsilon_{p}$, due to Coulomb interactions on this ligand.

\section{Derivation of the effective model}
\label{sec:procedure}
\noindent
This section provides a general description of our method. 
The $pp$ hopping delocalizes the single-hole intermediate state over the network of ligand ions.
First, we compute the band states formed by isolated $p$ orbitals of ligand ions.
The iodine lattice is defined by a unit cell containing two iodine ions 1 and 2 and lattice vectors $a_{1,2,3}$ as shown in Fig.~\ref{ch5:fig:CrI3lattice}{(a)}.
Diagonalizing the tight-binding Hamiltonian that includes $pp$ hopping and SOC yields Bloch states describing $p$ orbitals of I ions $V_{k,n}$ with wave vector $\bm k$ and band index $n$. The undistorted iodine sublattice has two I ions per unit cell, resulting in twelve bands shown in Fig.~\ref{ch5:fig:CrI3bands}(a).

The effective spin-spin Hamiltonian must contain all amplitudes of virtual transitions between two Cr sites.   
The effective amplitudes that describe these transitions can be factorized into hopping and spin parts. 
The hopping part is given by the effective $dd$ hopping amplitudes in second-order perturbation in the $dp$ hopping amplitudes:
\begin{equation}
\label{ch5:eq:Hdd1}
\begin{split}
&\left[t_{d_B, d_A}^{\sigma_B \sigma_A} \left(\bm X_B - \bm X_A \right)\right]_S=\sum_{\bm k,n}  \\
&\frac{
\bra{d_B, \bm X_B, \sigma_B}
H_{\rm TB}
\ket{V_{\bm k,n}}
\bra{V_{\bm k,n}} 
H_{\rm TB}
\ket{d_A, \bm X_A, \sigma_A}
}
{\left[\Delta_{\bm k,n}\right]_S},
\end{split}
\end{equation}
where an electron effectively hops from site $A$ to site $B$, with coordinates $\bm X_{A}$ and $\bm X_{B}$ via the band state $V_{\bm k,n}$ as schematically sketched in Fig.~\ref{ch5:fig:CrI3bands}{(b)}. $d_{A,B}$ denote the orbital state at a Cr site of the hopping electron with spin projection $\sigma_{A,B}$; $[\Delta_{\bm k,n}]_S$ is the energy gap of the single-hole excited state, depending on the band energy $\varepsilon_{k,n}$ and the total spin of four electrons in the intermediate state ($S=1$ or $S=2$), as described by the Kanamori Hamiltonian. 
Importantly, spin and orbital degrees of freedom are tightly entangled in $V_{\bm k,n}$ due to strong SOC at iodine ions, leading to effective $dd$ amplitudes with a spin flip (that is, $\sigma_A = -\sigma_B$).
The spin part is given by Clebsch-Gordan coefficients, 
\begin{equation}
 C^{S,M_B+\sigma_B}_{\frac{3}{2},M_B;\frac{1}{2},\sigma_B}
 C^{\frac{3}{2},M_A}_{1,M_A-\sigma_A;\frac{1}{2},\sigma_A}
 ,
\end{equation}
where $M_{A,B}$ are the total spin projections of three spins at sites $A,B$. They describe spin-wavefunction overlaps along the $dd$ transitions, and the projection in the final state on the manifold of Cr spin states with total spin 3/2.
We note that our $A$ and $B$ are as given in Fig.~\ref{ch5:fig:CrI3lattice}{(a)} for illustration, but all equations are written in a form that is applicable to arbitrary site positions.

Second, we calculate the amplitudes of the electron hopping between a ligand and a transition metal ion in an intermediate state. For this calculation, it is convenient to decompose the $p$ band wavefunctions into the atomic basis:
\begin{equation}
\ket{V_{\bm k,n}} = \sum_{a, \alpha, \sigma} c_{\bm k,n}^{a, \alpha, \sigma}
\ket{\bm k, a, p_\alpha, \sigma} ,
\end{equation}
where the plane waves in momentum space are given by
\begin{equation}
\label{ch5:eq:PinKspace}
\ket{\bm k, a, p_\alpha, \sigma}
= 
\frac{1}{\sqrt{N}}\sum_{i} e^{i \bm k \cdot \bm x_{i} }
\ket{\bm x_i, a, p_\alpha, \sigma},
\end{equation}
with $N$ being the total number of wave vectors in the Brillouin zone and $\bm x_i$ labeling the position of the unit cell of two iodine ions $a = 1,2$, as in Fig.~\ref{ch5:fig:CrI3lattice}{(a)}. 
For example, the amplitude of hopping to a $d_{xy}$ orbital is then given by
\begin{equation}
\begin{split}
&
\bra{d_{xy}, \bm X, \sigma}
H_{\rm TB}
\ket{V_{\bm k,n}} = t_{pd\pi}
\frac{e^{i \bm k \cdot \bm X} }{\sqrt{N}}\\&
\left(
c_{\bm k,n}^{1, x, \sigma} 
- e^{-i \bm k \cdot \left(\bm a_1 + \bm a_2\right)}
c_{\bm k,n}^{2, x, \sigma}
+
c_{\bm k,n}^{2, y, \sigma}
- e^{-i \bm k \cdot \left(\bm a_1 + \bm a_2\right)}
c_{\bm k,n}^{1, y, \sigma}
\right).
\end{split}
\end{equation}
Note that the Cr site position $\bm X$ can always be expressed in $\bm x_i$. The matrix elements that describe hopping to other orbitals and the energies of intermediate states can be found in Appendix~\ref{ap:hop_and_energies}.
Third, we obtain the effective $dd$ hopping amplitudes, as in Eq.\eqref{ch5:eq:Hdd1}, describing the long-range electron hopping between the $d$ orbitals of Cr ions. These three steps effectively integrate out $p$ orbitals, resulting in a $dd$ hopping model.

Finally, we map interactions between Cr spins in second-order perturbation theory in the $dd$ hopping amplitude to an effective Hamiltonian that describes spin-spin interactions. 
The spin Hamiltonian is expressed as:
\begin{equation}
\label{ch5:eq:effSpinModel}
\begin{split}
\mathcal{H}_{AB} =& J  (\bm S_A \cdot \bm S_B )    
+ K S_A^z S_B^z + \Gamma^{xy}\left( S_A^x S_B^y + S_A^y S_B^x \right)
\\&
+ \Gamma^{yz}\left( S_A^y S_B^z + S_A^z S_B^y \right)
+ \Gamma^{xz}\left( S_A^z S_B^x + S_A^x S_B^z \right)\\&
+ \bm D [ \bm S_A \times \bm S_B ]
,
\end{split}
\end{equation}
where $\bm{S}_{A,B}$ are the spin operators (total spin $S=3/2$), and $x,y,z$ are the global coordinates, taken throughout the paper as local coordinates of the Cr-I plaquette, as in Fig.~\ref{ch5:fig:CrI3lattice}{(a)}. The first term is the isotropic Heisenberg exchange, and the second is the out-of-plane exchange anisotropy. The three $\Gamma$ terms are the symmetric exchange anisotropy, and the last term is Dzyaloshinskii-Moriya anisotropic exchange. 
We neglect only the $\left(S_A^x S_B^x - S_A^y S_B^y\right)$ anisotropy compared to the most general $3\times3$ spin interaction matrix.
We perform the mapping by generating such spin model constants to match all the matrix elements that describe spin transitions between Cr spins in these two models, as detailed in Appendix~\ref{ap:JKGamma}.

Remarkably, we can use the same machinery to calculate single-ion anisotropy without complex adjustments. That is, by formally identifying two Cr sites, $\bm X_B = \bm X_A$, leads to effective $dd$ hoppings between orbitals $d_A$ and $d_B$ on the same Cr site, as detailed in Appendix~\ref{ap:SIA}. 
The quadratic single-ion anisotropy is given by
\begin{equation}
\mathcal{H}_{\rm SI} =  A_c \left( \bm S \cdot \hat {\bm c}\right)^2,
\end{equation}
with the out-of-plane direction $\hat{\bm c} = \frac{1}{\sqrt{3}}\left( \hat{\bm x} - \hat{\bm y} +\hat{\bm z} \right)$.

\section{Effective spin model}
\label{sec:sym}
\noindent
In order to clarify the origin of the anisotropy in the magnon spectrum, we introduce a spin model that includes nearest-neighbor and next-nearest-neighbor spin interactions and quadratic single-ion anisotropy.
%
We begin with the interactions between the nearest-neighbor Cr ions $A$ and $B$ in the $xy$ Cr-I plaquette. The monolayer symmetry, characterized by the dihedral point group $D_{3d}$, allows for the following spin exchange Hamiltonian:
\begin{equation}
\begin{split}
\mathcal{H}_{NN} = &
J_{1}  (\bm S_A \cdot \bm S_B ) +
 K_{1} S_A^z S_B^z
 + \Gamma_{1}^{xy}\left( S_A^x S_B^y + S_A^y S_B^x \right)
\\&
+ \Gamma_{1}^{xz}\left( S_A^z S_B^x + S_A^x S_B^z \right)
+ \Gamma_{1}^{yz}\left( S_A^y S_B^z + S_A^z S_B^y \right)
,
\end{split}
\end{equation}
In other neighboring $yz$- and $xz$-plaquettes, the three-fold symmetry $3_z$ ensures that the spin Hamiltonian has the same form as $\mathcal{H}_{NN}$ but in their local coordinates, which are $x' = z$, $y' = -x$, $z' = -y$ for $AB'$ and $x'' = -y$, $y'' = -z$, $z'' = x$ for $AB''$.
The $2_{[110]}$ symmetry restricts $\Gamma_{1}^{yz} = -\Gamma_{1}^{xz}$ and defines the $z$ direction of the Kitaev anisotropy; the inversion symmetry connecting the $A$ and $B$ sublattices disallows DMI.

As long as the exchange interactions are predominantly due to nearest-neighbor $pd$ hopping, we can consider the local symmetry of the ideal $xy$ plaquette, which is additionally symmetric under $2_{z}$ (see Fig.~\ref{ch5:fig:CrI3lattice}{(b)}), making $\Gamma_{1}^{yz}=0$.
Furthermore, the $\Gamma_{1}^{xy}$ term also vanishes due to $2'_{[110]}$ symmetry of the $dp$ model in the ideal $90^{\circ}$ exchange plaquette. This condition is lifted when trigonal distortions~\cite{Stavropoulos2021}, or the hopping between apex ligands~\cite{Barts2023} such as shown in Fig.~\ref{fig:hoppings}{(e)}, is included.


The exchange Hamiltonian for next-nearest-neighbor Cr ions $A$ and $C$ (as in Fig.~\ref{ch5:fig:CrI3lattice}{(a)}) is expressed as: 
\begin{equation}
\begin{split}
\mathcal{H}_{NNN} = &
J_{2}  (\bm S_A \cdot \bm S_C ) +
 K_{2} S_A^x S_C^x
+ \Gamma_{2}^{yz}\left( S_A^y S_C^z + S_A^z S_C^y \right)\\&
+ \Gamma_{2}^{xz}\left( S_A^z S_C^x + S_A^x S_C^z \right)
+ \Gamma_{2}^{xy}\left( S_A^x S_C^y + S_A^y S_C^x \right)\\&
+ \bm D [ \bm S_A \times \bm S_C ].
\end{split}
\end{equation}
Here, the $2_{[011]}$ symmetry implies that $\Gamma_{2}^{xy} = -\Gamma_{2}^{xz}$. 
The component of the DMI vector along the two-fold symmetry axis $2_{[011]}$ is zero, and the symmetry of the monolayer does not prohibit two other components. 

However, { \it all} components of the DMI vector are zero in our model.
This cancellation is enforced by an inversion center positioned at the center of the link connecting the sites $A$ and $C$ (see Fig.~\ref{ch5:fig:CrI3lattice}{(a)}), which prohibits DMI along this link.
This inversion operation does not belong to the symmetry group of the CrI$_3$ monolayer. However, it exists as a symmetry of a subsystem: the lattice of iodine ions and the individual Cr ions $A$ and $C$. Hence, it is the symmetry of our microscopic model because effective $dd$ interactions between these two ions are mediated only via the iodine states, while other Cr ions are not involved.

Remarkably, the $pp$ hopping also leads to quadratic single-ion anisotropy. It is not allowed by the cubic symmetry of the iodine octahedra coordinating individual Cr ions. However, introducing $pp$ hopping between ligand sites lowers the cubic symmetry to the dihedral symmetry of CrI$_3$ monolayer, even without symmetry-allowed lattice distortions. This symmetry-breaking mechanism leads to the quadratic single-ion anisotropy.

The symmetry of spin interactions is tightly bound to the symmetry of the underlying microscopic processes that mediate electron virtual exchange, being fundamental for our analysis. 
On the one hand, finite $A_c$ becomes allowed when $pp$ hoppings are included because new emergent virtual paths probe the ligand network with symmetry lower than that of the highly symmetric octahedral ligand environment. 
On the other hand, next-nearest-neighbor DMI is prohibited by the symmetry of the $dp$ model whereas the symmetry group of the monolayer has no such restriction. These arguments suggest the hierarchy of symmetries in the microscopic model that governs the hierarchy of magnetic interactions in the phenomenological model.

\section{Spin model parameters and magnon spectrum}

\begin{figure}
    \centering
    \includegraphics[width=1.0\linewidth]{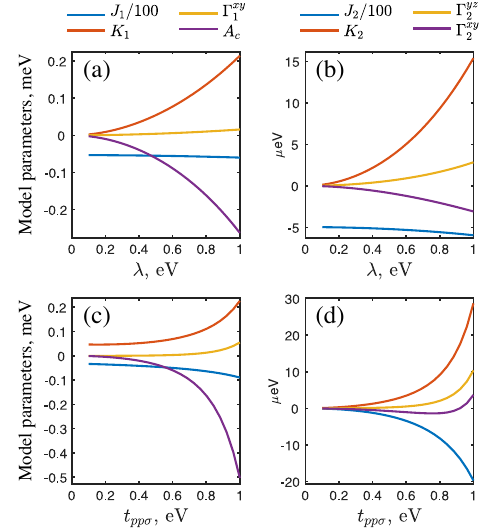}
    \caption{\label{ch5:fig:HamvsLambda} 
    Calculated spin model parameters.
    (a,b) SOC dependence of the single-ion anisotropy $A_c$ and $JK\Gamma$ parameters for (a) nearest- and {(b)}~next-nearest-neighbor Cr ions; $t_{pp\sigma}=0.7$~eV.
     (c,d) $pp$-hopping amplitude dependence of $A_c$ and $JK\Gamma$ for (c) nearest- and {(d)}~next-nearest-neighbor Cr ions; $\lambda=0.63$~eV.
	Other parameters are: 
	$t_{pp\pi}=t_{pp\sigma}/4$, $t_{pd\sigma}=1$~eV, $t_{pd\pi}=0.5$~eV 
	$U=3$~eV, $J_{\rm H} = 0.25 U$, $\Delta_c = 1.1$~eV and $U' = U - 2J_{\rm H}$. 
	}
     \end{figure}
\noindent
Finally, we calculate spin model parameters microscopically in monolayer ferromagnet CrI$_3$ when $pp$ hopping is included using the procedure developed in Sec.~\ref{sec:procedure}.
The resulting spin model parameters, depending on SOC- and $pp$-hopping amplitudes, are shown in Fig.~\ref{ch5:fig:HamvsLambda}. 
The nearest-neighbor and next-nearest-neighbor Heisenberg exchange coupling parameters $J_{1,2}$ remain relatively insensitive to $\lambda$. SOC gives rise to all anisotropy terms, as is evident in panels (a) and (b) where the $pp$ hopping amplitudes are kept finite and fixed. 
In panel~(a), we observe sizeable values of the nearest-neighbor antiferromagnetic Kitaev coupling ($K_1>0$) and the single-ion anisotropy ($A_c<0$), which always favors an out-of-plane spin arrangement. However, the nearest-neighbor $\Gamma_{1}^{xy}$ term remains small. 
In panel~(b), we find negligible anisotropy in the next-nearest-neighbor exchange interactions.
While SOC is usually assumed to be the main driver of magnetic anisotropy, 
the $t_{pp\sigma}$ dependencies of the spin model parameters shown in panels (c) and (d) nearly copy the SOC dependencies, highlighting the importance of $pp$ hopping for magnetic anisotropy in van der Waals materials.

This analysis suggests that the $J_{1,2,3} K_1 A_c$ model, which includes three Heisenberg exchange interactions, Kitaev nearest-neighbor interaction, and quadratic single-ion anisotropy, is the realistic low-energy model describing magnetism in CrI$_3$ monolayer. 
Using the microscopic Hubbard model parameters and hopping amplitudes from first-principles calculations, as introduced in Sec.~\ref{sec:micro}, the resulting spin model parameters are: 
    $J_1 = -2.12$~meV, 
$J_2 = -0.2$~meV,  
$J_3 = 0.005$~meV, 
$A_c = -0.1$~meV and $K_1=0.04$~meV, 
in good agreement with experimental values~\cite{Chen2021a}.

\begin{figure}[htb]
    \centering
\includegraphics[width=0.99\linewidth]{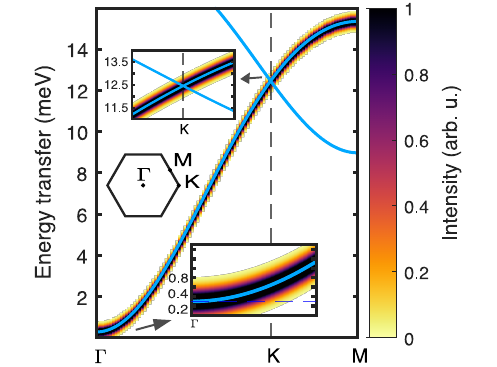}
    \caption{\label{fig:magnons} {Magnon energy spectrum} (blue lines) along the high symmetry lines 
    $\Gamma -K-M$ in the honeycomb Brillouin zone. It is obtained using spinW~\cite{Toth2014}. Color encodes the unpolarized neutron scattering cross-section. The insets show the spectrum details near the $\Gamma$ and $K$ points. The model parameters are: $J_1=-2.12$~meV, $J_2=-0.2$~meV, $J_3=0.005$~meV, $A_c = -0.1$~meV and $K_1=0.04$~meV,  and other are set to zero.
    }
\end{figure}

The resulting magnon spectrum shown in Fig.~\ref{fig:magnons} is consistent with neutron experiments on bulk samples, including the magnon bandwidth and the energy gap at the $\Gamma$ point~\cite{Chen2021a}. 
The zone-center energy gaps of the acoustic magnon ($\approx0.3$~meV) and the optical magnon ($\approx17$~meV) are in good agreement with Raman measurements for monolayer CrI$_3$~\cite{Cenker2021}.
The gap at the $\Gamma$ point originates solely from single-ion anisotropy, as the contribution from $\Gamma_{1}^{xy}$ is tiny due to its small magnitude, and the Kitaev term does not contribute to it at all.

However, the calculated gap at the $K$ point is very small. The next-nearest-neighbor DMI is absent in our model due to the hidden inversion symmetry of the sublattice of iodines; the Kitaev interaction contributes only weakly to the Dirac gap, with the energy gap scaling as $\left(K_1\right)^2/J_1$; and other anisotropy sources are too weak in magnitude. Therefore, our model cannot explain the experimentally reported large Dirac gap.

\section{Conclusions and outlook}
\noindent
We discussed anisotropic exchange interactions in materials with strong SOC on ligand sites within the extended Hubbard model that includes $pp$ hopping. 
This model allows for the calculation of long-range exchange interactions. We proposed an effective method of doing it, which we applied to calculate the anisotropic interactions between the nearest- and next-nearest-neighbor spins in the monolayer CrI$_3$. 
Using model parameters from first-principles calculations, we obtain the spin-wave spectrum in good agreement with experimental data everywhere except in the vicinity of Dirac points, where the energy gap is much smaller than the experimental value. 
Our calculations show that both Kitaev and next-nearest-neighbor Dzyaloshinskii-Moriya interactions are unlikely sources of the large gap. 
Issues with the Kitaev scenario were previously understood~\cite{Chen2021a}. We further find that the next-nearest-neighbor DMI vanishes if trigonal distortions of the CrI$_3$ lattice are neglected.
Although recent studies suggest that the strong hybridization between magnons and phonons at the Dirac points may enhance the gap~\cite{Delugas2023}, a relatively small distortion is unlikely to induce the large gap.

We also demonstrated that the ``dressing'' of Cr spins by states with ligand holes induces an effective SOC on the transition metal sites. This produces the uniaxial anisotropy that is second order in the Cr spin operators on the same site. 
This single-ion anisotropy is the main source of the magnon energy gap at the Brillouin zone center, preventing the destruction of the ferromagnetic order in the monolayer CrI$_3$ by spin fluctuations. 
The second-order anisotropy requires low symmetry of the transition metal sites, which in our model is a result of the delocalization of holes over the ligand network due to the $pp$ hopping.
The resulting uniaxial anisotropy is comparable to that originating from trigonal distortions~\cite{Stavropoulos2021}.

Our method can be extended to study interlayer interactions in twisted bilayer CrI$_3$, which can lead to non-collinear states~\cite{Tong2018, Hejazi2020} and coexisting antiferromagnetic and ferromagnetic domains~\cite{Xu2021}.
It can also be applied to other van der Waals ferromagnets with strong SOC on ligand sites, such as CrGeTe$_3$/CrSiTe$_3$~\cite{Gong2017} with Te ligand ions, where a similar magnon gap hierarchy has been reported~\cite{Zhu2021,Chen2022}, and can be extended to include intermediate states with two ligand holes.

Understanding longer-range interactions between spins is also important because they frustrate the uniform ferromagnetic state and can result in spiral magnetic orders~\cite{Okubo2012,Leonov2015,Kurumaji2019}, as it happens in FeCl$_3$ and a number of transition metal dihalides with the relatively strong third-nearest-neighbor exchange~\cite{McGuire2017, Amoroso2020, Song2022}. 
Magnetic anisotropy in van der Waals materials can stabilize multiple periodic magnetic states and isolated skyrmions~\cite{Behera2019,Park2019} and enable the electric control of 
magnetism~\cite{Gong2019,Huang2022}, magnetic topological defects~\cite{Mostovoy2023} and potentially magnon edge states.

\section*{Acknowledgments}
\noindent
EB thanks Francesco Foggetti, Mihir Date, Jagoda S\l awi\'{n}ska, and Ming-Chun Jiang for fruitful discussions. EB and MM acknowledge Vrije FOM-programma `Skyrmionics'. P.B. would like to thank Dr. J. S\l awi\'{n}ska for her kind invitation to visit Zernike Institute for Advanced Materials, where he had
 the opportunity to take part in this project. We acknowledge the CINECA award under the ISCRA initiative Grant No.
HP10CWPBNU for the availability of high-performance computing resources
and support. 

\appendix

\section{Computational details and derivation of hopping integrals from first principles}\label{app:dft}

\noindent
We considered an ideal CrI$_3$ monolayer with the in-plane lattice constant fixed to the experimental value of $a_0 = 6.87~$\AA \cite{McGuire2015} and a vertical spacing  of 22~\AA~ between periodic replicas of the monolayer. The appropriate layer group for the monolayer is $p\bar{3}1m$, belonging to the trigonal $D_{3d}$ crystal class, where chromium atoms occupy the Wyckoff position $2c$ with fractional coordinates $(\frac{1}{3},\frac{2}{3},0)$ and the octahedrally coordinated iodine atoms occupy the Wyckoff position $6k$ with fractional coordinates $(x,0,z)$. Experimentally, CrI$_3$ undergoes a trigonal distortion along the out-of-plane $c$ axis that keeps the crystal symmetry but causes a deformation of both the CrI$_6$ octahedra and of the Cr-I plaquette, thus lowering the local symmetries. We remove such trigonal distortion by choosing appropriately the iodine Wyckoff coordinates in order to enforce octahedral symmetry on Cr and ideal square Cr-I plaquettes. Non-relativistic spin-unpolarized band structure has been calculated using ultrasoft potentials generated with the Rappe-Rabe-Kaxiras-Joannopoulos method\cite{rrkjus} and PBE functional within the generalized gradient approximation\cite{PBE} as implemented in Quantum ESPRESSO\cite{qe}. A 24$\times$24$\times$1 Monkhorst-Pack grid and cold smearing\cite{qe_smearing} of 0.05 Ry has been adopted for Brillouin-zone integration, using a Coulomb cutoff along the perpendicular direction to simulate an isolated CrI$_3$ monolayer\cite{qe_2dcutoff}. We resort to the  Wannier90 package\cite{w90} to project the band structure on maximally localized Wannier functions with orbital character $\{d_{xy},d_{yz},d_{xz},d_{x^2-y^2},d_{3z^2-r^2}\}$ for chromium and $\{p_x, p_y, p_z\}$ for iodine as defined in the global cartesian reference frame shown in Fig.~\ref{ch5:fig:CrI3lattice}. 

\begin{table}[htb]
\centering
\begin{tabular}{lccc}
\hline
      & $p_x$ & $p_y$ & $p_z$ \\
\hline
$d_{yz}$      & 0.005  & 0.000  & 0.000 \\
$d_{xz}$      & 0.003  & -0.003 & -0.477 \\
$d_{xy}$      & 0.013  & -0.468 & 0.145 \\
$d_{z^2}$     & -0.496 & -0.019 & 0.000 \\
$d_{x^2-y^2}$ & 0.854  & 0.021  & 0.005 \\
\hline
\end{tabular}
\caption{Hopping parameters expressed in eV between Cr$_A$-$d$ states and I$_1$-$p$ states in the plaquette shown in Fig.~\ref{ch5:fig:CrI3lattice}. Diagonal energies for Cr-$d$ states are $\varepsilon_{t_{2g}} = -5.611$ eV and $\varepsilon_{e_g} = -5.240$ eV; diagonal energies for I-$p_x, p_y$ states lying in the plaquette plane and I-$p_z$ states perpendicular to it are $\varepsilon_{p_\parallel} = -7.662$ eV and $\varepsilon_{p_\perp} = -7.013$ eV.}
\label{tab:pd_wannier}
\end{table}

\begin{table}[htb]
\centering
\begin{tabular}{lccc}
\hline
      & $p_x$ & $p_y$ & $p_z$ \\
\hline
$p_x$ & 0.277  & -0.422 & 0.003 \\
$p_y$ & -0.422 & 0.277  & -0.003 \\
$p_z$ & 0.003  & -0.003 & -0.061 \\
\hline
\end{tabular}
\caption{Hopping parameters expressed in eV between I$_1$-$p$ and I$_2$-$p$ states in the plaquette shown in Fig.~\ref{ch5:fig:CrI3lattice}.}
\label{tab:pp_wannier}
\end{table}

\begin{table}[htb]
\centering
\begin{tabular}{lccccc}
\hline
      & $d_{yz}$ & $d_{xz}$ & $d_{xy}$ & $d_{z^2}$ & $d_{x^2-y^2}$ \\
\hline
$d_{yz}$      & 0.038 & 0.015 & 0      & 0      & 0 \\
$d_{xz}$      & 0.015 & 0.038 & 0      & 0      & 0 \\
$d_{xy}$      & 0     & 0     & -0.142 & 0.009  & 0 \\
$d_{z^2}$     & 0     & 0     & 0.009  & 0.011  & 0 \\
$d_{x^2-y^2}$ & 0     & 0     & 0      & 0      & -0.097 \\
\hline
\end{tabular}
\caption{Direct hopping parameters (in eV) between Cr$_A$-$d$ and Cr$_B$-$d$ states.}
\label{tab:dd_wannier}
\end{table}

We report here the relevant hopping integrals between $pd$ and $pp$ Wannier states on the Cr-I plaquette formed by atoms labeled $A$, $B$ (chromium $d$ states) and 1,2 (iodine $p$ states) in Fig.~\ref{ch5:fig:CrI3lattice}. Slater-Koster parameters in the ideal plaquette can be estimated using the following relations for hopping matrix elements $E({\alpha,\beta})$, where greek indices denote Wannier states. For $pd$ hopping matrix elements listed in Table~\ref{tab:pd_wannier}, one has:
\begin{eqnarray}
t_0 &=& E(z,zx) \equiv E(y,xy) = -V_{pd\pi}, \nonumber\\
t_1 &=& E(x,x^2-y^2) = -\frac{\sqrt{3}}{2}V_{pd\sigma}, \nonumber\\
t_2 &=& E(x,z^2)  = \frac{1}{2}V_{pd\sigma},
\end{eqnarray}
and for $pp$ hopping matrix elements listed in Table \ref{tab:pp_wannier}:
\begin{eqnarray}\label{eq:pp-in}
t_{p1}&=& E(x,x) \equiv E(y,y) = \frac{V_{pp\sigma}+V_{pp\pi}}{2},\nonumber\\
t_{p2}&=& E(x,y) = \frac{-V_{pp\sigma}+V_{pp\pi}}{2},\nonumber\\
t_{p3} &=& E(z,z)= V_{pp\pi}.
\end{eqnarray}
For completeness, we also report the relations for direct $dd$ hopping terms, given in Table \ref{tab:dd_wannier} :
\begin{eqnarray}
{t}_{d0} &=& E(xy,z^2) = \frac{\sqrt{3}}{4}\left(-V_{dd\sigma}+V_{dd\delta}\right) \nonumber\\
t_{d1} &=& E(yz,yz) \equiv E(zx,zx) = \frac{V_{dd\pi}+V_{dd\delta}}{2},\nonumber\\
t_{d2} &=& E(yz,zx) \equiv E(zx,yx) = \frac{V_{dd\pi}-V_{dd\delta}}{2},\nonumber\\
t_{d3} &=& E(xy,xy) = \frac{3 V_{dd\sigma}+V_{dd\delta}}{4}.
\end{eqnarray}

\section{Hopping amplitudes and energies of intermediate states}
\label{ap:hop_and_energies}
\noindent
Here hopping amplitudes and the energies of all intermediate states are listed.
The Hamiltonian for hopping between $d_{yz, A}$ and $p_{z,1}$ orbitals is
\begin{equation}
-t_{pd\pi} \biggl( \ket{d_{yz,A}}\bra{p_{z,1}} + \ket{p_{z,1}}\bra{d_{yz,A}} \biggr).
\end{equation}
The amplitudes are chosen positive, so that $t_{pd\pi} = V_{pd\pi}>0$ and $t_{pd\sigma}= - V_{pd\sigma}>0$, and $t_{pp\pi} = V_{pp\pi}>0$ and $t_{pp\sigma} = -V_{pp\sigma}>0$ in the Slater-Koster notation~\cite{Slater1954}.
The Kanamori Hamiltonian is 
 \begin{equation}
\begin{split}
H_{{\rm ee}} =& U\sum_\alpha n_{\alpha\uparrow} n_{\alpha\downarrow} 
+
U^{'}\sum_{\alpha \neq \beta,\sigma,\sigma^{'}} n_{\alpha\sigma} n_{\beta\sigma'}\\ 
& - \frac{J_{\rm H}}{2}\sum_{\alpha \neq \beta,\sigma,\sigma^{'}} d_{\alpha\sigma}^{\dagger}
d_{\beta\sigma'}^{\dagger}
d_{\beta\sigma} 
d_{\alpha\sigma'}
\\
&
 + J_{\rm H}\sum_{\alpha \neq \beta} d_{\alpha\uparrow}^{\dagger}
 d_{\alpha\downarrow}^{\dagger}
d_{\beta\downarrow} d_{\beta\uparrow},
\end{split}
\end{equation}
where $d_{\alpha\sigma}$ annihilates electron with spin projection $\sigma$ in orbital state $\alpha$ on a Cr site, and $n_{\alpha\sigma} = d_{\alpha\sigma}^\dagger d_{\alpha\sigma}$.
$U$ is the Coulomb repulsion between two electrons occupying the same $d$ orbital, $U^{'}$ is the energy of two electrons at different orbitals with antiparallel spins and $(U^{'} - J_{\rm H})$ is the Coulomb energy for electrons with parallel spins.
The Hund's rule coupling gives rise to off-diagonal matrix elements in the Hamiltonian describing electrons on $d$-sites, i.e., electron pair hopping from one $d$ orbital to another with the amplitude $J_{\rm H}$ and flipping of spins of two electrons with opposite spin projections at two $d$ orbitals with the amplitude $(-J_{\rm H})$. 

For hopping between $t_{2g}$ orbitals on two Cr sites, the intermediate state energies are
\begin{equation}
\label{eq:DeltaU1}
\Delta_{S=1}^{t_{2g}} = \varepsilon_{t_{2g}} - \varepsilon_{p} + U +2 U^{'} ,
\qquad 
U_{S=1}^{t_{2g}} = U +2 J_{\rm H},
\end{equation}
where $\Delta_S$ is the energy of the state with a single hole at a ligand site and the total spin $S$ at the metal site with four electrons, and $U_S$ is the energy of the state with four electrons on one transition metal site and two electrons on the other transition metal site.
For $e_{g}-t_{2g}$ hopping processes with the total spin $S=1,2$ of four electrons on one Cr ion and two electrons on the second Cr ions in the intermediate state, the energies are:
\begin{equation}
\Delta_{S=1}^{e_{g}} = \varepsilon_{e_g} - \varepsilon_{p} +3 U^{'} + J_{\rm H},
\quad 
U_{S=1}^{e_{g}} = \Delta_c + U^{'} + 3 J_{\rm H},
\end{equation}
\begin{equation}
\Delta_{S=2}^{e_{g}} = 
\Delta_{S=1}^{e_{g}}
 -4 J_{\rm H},
\quad
U_{S=2}^{e_{g}} = U_{S=1}^{e_{g}} -4J_{\rm H}.
\end{equation}
The energies relevant for the $e_{g}-t_{2g}$ hopping processes with the total spin $S=1/2,3/2$ of three electrons on a Cr ion in the first excited states are:
\begin{equation}
U_{S=3/2}^{e_{g}} = \varepsilon_{e_g} - \varepsilon_{t_{2g}} , 
\quad 
U_{S=1/2}^{e_{g}} = U_{S=3/2}^{e_{g}} +3 J_{\rm H}.
\end{equation}

The amplitudes of hopping to a $d$ orbital from $p$ orbitals of nearest iodine ions are listed below:
\begin{widetext}
\begin{equation}
\begin{split}
\bra{d_{xy}, \bm X, \sigma}
H_{\rm TB}
\ket{V_{\bm k,n}} =& t_{pd\pi}
\frac{e^{i \bm k \cdot \bm X} }{\sqrt{N}}
\left(
c_{\bm k,n}^{1, x, \sigma} 
- e^{-i \bm k \cdot \left(\bm a_1 + \bm a_2\right)}
c_{\bm k,n}^{2, x, \sigma}
+
c_{\bm k,n}^{2, y, \sigma}
- e^{-i \bm k \cdot \left(\bm a_1 + \bm a_2\right)}
c_{\bm k,n}^{1, y, \sigma}
\right), \\
\bra{d_{yz}, \bm X, \sigma}
H_{\rm TB}
\ket{V_{\bm k,n}} =& t_{pd\pi}
\frac{e^{i \bm k \cdot \bm X} }{\sqrt{N}}
\left(
c_{\bm k,n}^{1, z, \sigma} 
- e^{-i \bm k \cdot \left(\bm a_1 + \bm a_2\right)}
c_{\bm k,n}^{2, z, \sigma}
+
e^{-i \bm k \cdot \bm a_1 }
c_{\bm k,n}^{2, y, \sigma}
- e^{-i \bm k \cdot \bm a_2 }
c_{\bm k,n}^{1, y, \sigma}
\right), \\
%
\bra{d_{xz}, \bm X, \sigma}
H_{\rm TB}
\ket{V_{\bm k,n}} =& t_{pd\pi}
\frac{e^{i \bm k \cdot \bm X} }{\sqrt{N}}
\left(
e^{-i \bm k \cdot \bm a_1 }
c_{\bm k,n}^{2, x, \sigma} 
- e^{-i \bm k \cdot \bm a_2}
c_{\bm k,n}^{1, x, \sigma}
+
c_{\bm k,n}^{2, z, \sigma}
- e^{-i \bm k \cdot \left(\bm a_1 + \bm a_2\right)}
c_{\bm k,n}^{1, z, \sigma}
\right),\\
\end{split}
\end{equation}
\begin{equation}
\begin{split}
\bra{d_{x^2 - y^2}, \bm X, \sigma}
H_{\rm TB}
\ket{V_{\bm k,n}} =& \frac{\sqrt{3}}{2}t_{pd\sigma}
\frac{e^{i \bm k \cdot \bm X} }{\sqrt{N}}
\left(
c_{\bm k,n}^{1, y, \sigma}
- e^{-i \bm k \cdot \left(\bm a_1 + \bm a_2\right)}
c_{\bm k,n}^{2, y, \sigma}
-
c_{\bm k,n}^{2, x, \sigma}
+ e^{-i \bm k \cdot \left(\bm a_1 + \bm a_2\right)}
c_{\bm k,n}^{1, x, \sigma}
\right),
\\
\bra{d_{3z^2 - r^2}, \bm X, \sigma}
H_{\rm TB}
\ket{V_{\bm k,n}} =& \frac{1}{2}t_{pd\sigma}
\frac{e^{i \bm k \cdot \bm X} }{\sqrt{N}}
(
c_{\bm k,n}^{1, y, \sigma}
- e^{-i \bm k \cdot \left(\bm a_1 + \bm a_2\right)}
c_{\bm k,n}^{2, y, \sigma}
+
c_{\bm k,n}^{2, x, \sigma}
- e^{-i \bm k \cdot \left(\bm a_1 + \bm a_2\right)}
c_{\bm k,n}^{1, x, \sigma}\\& 
-2 
e^{-i \bm k \cdot \bm a_1}
c_{\bm k,n}^{2, z, \sigma}
+2 
e^{-i \bm k \cdot \bm a_2}
c_{\bm k,n}^{1, z, \sigma}
).
\end{split}
\end{equation}
\end{widetext}

\section{Spin model parameters}
\label{ap:JKGamma}
\noindent
{\bf $JK\Gamma$ parameters.}
Hopping processes that lead to the $JK\Gamma$ model are discussed in detail in Ref.~\cite{Barts2023}, and their analytical expressions when $pp$ hopping and distortions are excluded, match those in Ref.~\cite{Stavropoulos2021}. 

The isotropic Heisenberg exchange coupling:
\begin{equation}
\label{eq:J_Heis}
\begin{split}
&J = 
\frac{2}{9 U_1^{t_{2g}}}
 \sum_{\substack{d_B \in t_{2g}\\d_A \in t_{2g}}}
\left[
 \abs{t_{d_B, d_A}^{ \uparrow \uparrow}}^2
 +
 \abs{t_{d_B, d_A}^{ \downarrow \downarrow}}^2
 \right]_{S=1}
  \\&
  +
 \sum_{S=1,2} \frac{(-1)^{S+1}}{6 U_S^{e_{g}}}
  \left[
 \sum_{\substack{d_B \in e_g\\d_A \in t_{2g}}}
 \abs{t_{d_B, d_A}^{ \uparrow \uparrow}}^2
 +
 \sum_{\substack{d_B \in t_{2g} \\d_A \in e_{g} }}
 \abs{t_{d_B, d_A}^{ \downarrow \downarrow}}^2
 \right]_S,
 \end{split}
\end{equation}
where the $dd$ hoppings dependence on $(\bm X_B - \bm X_A)$ is implicitly assumed. The first line corresponds to hoppings between $t_{2g}$ orbitals, shown in Fig.~\ref{fig:hoppings}{(a)}. %
The second line corresponds to hoppings between $t_{2g}$ and $e_g$ orbitals, shown in Fig.~\ref{fig:hoppings}{(b)}.

Hopping processes where the electron spin flips twice in two intermediate states at a ligand site, as exemplified in Figs.~\ref{fig:hoppings}{(c),(d)}, lead to the Kitaev exchange coupling:
\begin{equation}
\begin{split}
&K = 
-\frac{2}{9 U_1^{t_{2g}}}
 \sum_{\substack{d_B \in t_{2g}\\d_A \in t_{2g}}}
\left[
 \abs{t_{d_B, d_A}^{ \downarrow \uparrow}}^2
 +
 \abs{t_{d_B, d_A}^{ \uparrow \downarrow}}^2
 \right]_{S=1}
  \\&
  -
 \sum_{S=1,2} \frac{(-1)^{S+1}}{6 U_S^{e_{g}}}
  \left[
 \sum_{\substack{d_B \in e_g\\d_A \in t_{2g}}}
 \abs{t_{d_B, d_A}^{ \downarrow \uparrow}}^2
 +
 \sum_{\substack{d_B \in t_{2g} \\d_A \in e_{g} }}
 \abs{t_{d_B, d_A}^{ \uparrow \downarrow}}^2
 \right]_S.
 \end{split}
\end{equation}
The prefactors in this expression resemble those in Eq.~\eqref{eq:J_Heis}, but the overall sign is opposite, suggesting antiferromagnetic Kitaev interaction (i.e., $K>0$) when the $S=2$ hopping channel is dominating.

Similar processes but with an additional spin flip at the Cr site due to Hund's rule coupling lead to the $xy$ symmetric off-diagonal exchange coupling:
\begin{equation}
\begin{split}
&\Gamma^{xy} = {\rm Im} \sum_{S=1,2} \frac{(-1)^{S+1}}{6 U_S^{e_{g}}} 
  \\&
 \left[
 \sum_{\substack{d_B \in e_g\\d_A \in t_{2g}}}
 t_{d_B, d_A}^{ \downarrow \uparrow} \left(t_{d_B, d_A}^{ \uparrow \downarrow}\right)^*
 +
 \sum_{\substack{d_B \in t_{2g} \\d_A \in e_{g} }}
 \left(t_{d_B, d_A}^{ \uparrow \downarrow}\right)^*
 t_{d_B, d_A}^{ \downarrow \uparrow} 
 \right]_S.
 \end{split}
\end{equation}

\noindent
{\bf DMI, $\Gamma^{yz}$, and $\Gamma^{xz}$ parameters.}
The $z$ component of the DMI vector comes from hopping processes with a single spin flip at a Cr site due to Hund's rule:
\begin{equation}
\begin{split}
&D_{z} = {\rm Im} \sum_{S=1,2} \frac{(-1)^{S}}{6 U_S^{e_{g}}}   
\\&
 \left[
 \sum_{\substack{d_B \in e_g\\d_A \in t_{2g}}}
 t_{d_B, d_A}^{ \uparrow \uparrow} \left(t_{d_B, d_A}^{ \downarrow \downarrow}\right)^*
 +
 \sum_{\substack{d_B \in t_{2g} \\d_A \in e_{g} }}
 \left(t_{d_B, d_A}^{ \downarrow \downarrow}\right)^*
 t_{d_B, d_A}^{ \uparrow \uparrow} 
 \right]_S.
 \end{split}
\end{equation}
It is obtained by calculating the imaginary part of the matrix element ($M_{AB}^{-+}$) for the process with the interchange of the spin projections: $M_A' = M_A - 2\sigma$ and $M_B' = M_B + 2\sigma$, where $M_{A,B}'$ are the total spin projections in the final state.

To calculate $D_x$ and $D_y$, as well as the $\Gamma^{yz}$ and $\Gamma^{xz}$, we consider processes that lower the spin projection only on one of the Cr sites. 
By calculating their matrix elements in the effective spin model, we obtain the following set of expressions for exchange coupling constants:  
\begin{equation}
\begin{split}
D_x &=\frac{2}{3\sqrt{3}} {\rm Im}\left( M_{AB}^{-z} - M_{AB}^{z-} \right),
\\
\Gamma^{yz} &=\frac{2}{3\sqrt{3}} {\rm Im}\left( M_{AB}^{-z} +M_{AB}^{z-} \right),
\end{split}
\end{equation}
and 
\begin{equation}
\begin{split}
D_y &=-\frac{2}{3\sqrt{3}} {\rm Re}\left( M_{AB}^{-z} - M_{AB}^{z-} \right),
\\
\Gamma^{xz} &=\frac{2}{3\sqrt{3}} {\rm Re}\left( M_{AB}^{-z} +M_{AB}^{z-} \right).
\end{split}
\end{equation}
Here, $M_{AB}^{-z} $ is the matrix element in the effective $dd$ model that describes the processes resulting in lower spin projection at site $A$, $M_A' = M_A - 2\sigma$ and $M_B' = M_B$. Analogously, $M_{AB}^{z-} $ is for $M_A' = M_A$ and $M_B' = M_B - 2\sigma$. 
We will describe these processes in more detail below.

The first matrix element can be split into three, 
$M_{AB}^{z-} = M_{AB,1}^{z-} + M_{AB,2}^{z-} + M_{AB,3}^{z-}$. 
In the first term, 
\begin{equation}
\begin{split}
M_{AB,1}^{z-} =\frac{1}{\sqrt{3}}
 &\sum_{S=1,2} \frac{3(-1)^{S+1}}{4  U_S^{e_{g}}}  \\&
  \left[
 \sum_{\substack{d_B \in e_g\\d_A \in t_{2g}}}
 t_{d_B, d_A}^{ \downarrow \uparrow} \left(t_{d_B, d_A}^{ \uparrow \uparrow}\right)^* 
 \right]_S,
 \end{split}
\end{equation}
the spin-up electron hops from a $t_{2g}$ orbital at site $A$ to an $e_g$ orbital at site $B$ with a spin flip and returns to site $A$ in the spin-up state after Hund's rule spin exchange with one of the three electrons at site $B$. 
In the second term,
\begin{equation}
\begin{split}
M_{AB,2}^{z-} =
- \frac{1}{\sqrt{3}}
&\sum_{S=1,2}
 \frac{\left(3 \delta_{S,1} + \delta_{S,2} \right)}{4 U_S^{e_{g}} }\\&
 \left[
 \sum_{\substack{d_B \in t_{2g} \\d_A \in e_g }}
 \left(t_{d_B, d_A}^{ \uparrow \downarrow }\right)^* t_{d_B, d_A}^{ \downarrow \downarrow} 
 \right]_S ,
 \end{split}
\end{equation}
the spin up electron hops from a $t_{2g}$ orbital at site $B$ to an $e_g$ orbital at site $A$ with a spin flip and returns back to site $B$ in the spin down state. $\delta_{S,S'}$ is the Kronecker delta function. 
A similar process with spin-flip occurring on the return of the electron to site $B$ leads to 
\begin{equation}
M_{AB,3}^{z-} =
- 
\frac{1}{\sqrt{3} U_2^{e_{g}} } 
\left[
 \sum_{\substack{d_B \in t_{2g} \\d_A \in e_g }}
 \left(t_{d_B, d_A}^{ \uparrow \uparrow }\right)^* t_{d_B, d_A}^{ \downarrow \uparrow} 
 \right]_{S=2}  .
\end{equation}

The matrix element of the processes that lower the spin projection only on the site $A$ is obtained by analogy,  
$M_{AB}^{-z} = M_{AB,1}^{-z} + M_{AB,2}^{-z} + M_{AB,3}^{-z}$:
\begin{equation}
\begin{split}
M_{AB,1}^{-z} =\frac{1}{\sqrt{3}}
 &\sum_{S=1,2} \frac{3(-1)^{S+1}}{4  U_S^{e_{g}}}  \\&
  \left[
 \sum_{\substack{d_B \in t_{2g} \\d_A \in e_g }}
 \left(t_{d_B, d_A}^{ \uparrow \downarrow }\right)^* t_{d_B, d_A}^{ \uparrow \uparrow} 
 \right]_S ,
 \end{split}
\end{equation}
when an electron hops from site $B$, and the matrix elements when an electron hops from site $A$ are
\begin{equation}
\begin{split}
M_{AB,2}^{-z} =
- \frac{1}{\sqrt{3}}
&\sum_{S=1,2}
  \frac{\left(3 \delta_{S,1} + \delta_{S,2} \right)}{4 U_S^{e_{g}} }\\&
 \left[
 \sum_{\substack{d_B \in e_g\\d_A \in t_{2g}}}
 t_{d_B, d_A}^{ \downarrow \uparrow} \left(t_{d_B, d_A}^{ \downarrow \downarrow}\right)^* 
 \right]_S ,
 \end{split}
\end{equation}
and
\begin{equation}
M_{AB,3}^{-z} =
- 
\frac{1}{\sqrt{3} U_2^{e_{g}} } 
\left[
 \sum_{\substack{d_B \in e_g\\d_A \in t_{2g}}}
 t_{d_B, d_A}^{ \uparrow \uparrow} \left(t_{d_B, d_A}^{ \uparrow \downarrow}\right)^* 
 \right]_{S=2} .
 \end{equation}
We found that contributions to DMI constants from $t_{2g}-t_{2g}$ hopping processes are negligible, so they are not shown here.
This finalizes the calculation of the $D_x$, $D_y$, $\Gamma^{yz}$ and $\Gamma^{xz}$ off-diagonal and other exchange coupling constants.

\section{Single-ion anisotropy}
\label{ap:SIA}
\noindent
The quadratic single-ion anisotropy is obtained by calculating the matrix element of lowering the spin projection on a Cr site,
$M' = M - 2\sigma$:
\begin{equation}
A_c = \frac{\sqrt{3}}{2} \left(1 + i \right) M_{\rm SI}^{-}.
\end{equation}
We split the matrix element from the effective single-ion $dd$ model into four different contributions, 
$M_{\rm SI}^{-} = M_{\rm SI, 1}^{-} + M_{\rm SI,2}^{-} + M_{\rm SI, 3}^{-} + M_{\rm SI, 4}^{-}$. 

A complexity arises compared to the calculations in the previous sections because total spin 3/2 and 1/2 are now reachable in an intermediate state with a single hole on a $t_{2g}$ orbital and an occupied $e_g$ orbital.
The processes described by the matrix elements $M_{\rm SI, 1}^{-}$ and $M_{\rm SI, 2}^{-}$ involve only the $J=3/2$ state, while it is the $J=1/2$ for $M_{\rm SI, 3}^{-}$ and $M_{\rm SI, 4}^{-}$. 
The first matrix element,
\begin{equation}
\begin{split}
M_{\rm SI,1}^{-} =&
-\frac{1}{\sqrt{3}U_{3/2}^{e_g}}
\sum_{\substack{d_B \in e_g \\d_A \in t_{2g}}} 
\left(t_{d_B, d_A}^{ \uparrow \uparrow}\right)_{S=2} 
\\&
\left[
\frac{3}{4}\left(t_{d_B, d_A}^{ \uparrow \downarrow}\right)^*_{S=2} 
+
\frac{1}{4}\left(t_{d_B, d_A}^{ \uparrow \downarrow}\right)^*_{S=1}
\right],
\end{split}
\end{equation}
describes spin up electron that hops from a ligand orbital to an $e_g$ orbital on the Cr site. Next, one of the three $t_{2g}$ electrons hops back to the iodine lattice. A spin-up electron fills this $t_{2g}$ hole again. Note that we need to project the Cr four-electron spin state onto eigenstates of their total spin operator because only those have well-defined pair-wise Coulomb interaction energy in our model. In the final state, the spin state of the three electrons is projected onto the state with total spin $3/2$ and spin projection of $1/2$. 

The contributions from the other three matrix elements to $A_c$ are much smaller than $M_{\rm SI,1}^{-}$ as they describe processes in which electron first hops to an $e_g$ orbital in the spin down state, unfavored by Hund's rule coupling. The second process gives
\begin{equation}
\begin{split}
M_{\rm SI,2}^{-} =&
-\frac{1}{4\sqrt{3}U_{3/2}^{e_g}}
\sum_{\substack{d_B \in e_g \\d_A \in t_{2g}}} 
\left[
\left(t_{d_B, d_A}^{ \downarrow \downarrow}\right)_{S=2} 
-
\left(t_{d_B, d_A}^{ \downarrow \downarrow}\right)_{S=1} 
\right]\\&
\left[
\frac{3}{4}\left(t_{d_B, d_A}^{ \uparrow \downarrow}\right)^*_{S=2} 
+
\frac{1}{4}\left(t_{d_B, d_A}^{ \uparrow \downarrow}\right)^*_{S=1}
\right].
\end{split}
\end{equation}
The state with the spin down $e_g$ electron is projected on eigenstates of the total spin of four electrons. As on of the three $t_{2g}$ electrons with spin down hops away to a ligand ion, we end up in the same three-electron excited state as in the calculation of the matrix element $M_{\rm SI,1}^{-} $. With an additional factor of $1/\sqrt{3}$, due to the overlap between the spin states, 
$\bra{\frac{3}{2},\frac{1}{2}}\ket{\downarrow} \ket{1,1} = 1/\sqrt{3}$.

The second option is to annihilate the spin up $t_{2g}$ electron leading to the factor $\bra{\frac{3}{2},\frac{1}{2}}\ket{\uparrow} \ket{1,0} = \sqrt{2/3}$. The resulting excited spin state is projected on the eigenstate of the total spin, using the wave function decomposition for two $t_{2g}$ and one $e_{g}$ electron with the total spin projection $1/2$.
This decomposition leads to two independent contributions to the matrix elements, corresponding to  
$M_{\rm SI,3}^{-}= M_{\rm SI,3a}^{-} + M_{\rm SI,3b}^{-}$. 
$M_{\rm SI,3a}^{-}$ is expressed as
\begin{widetext}
\begin{equation}
\label{eq:ch5:MAA3a}
\begin{split}
M_{\rm SI,3a}^{-} = 
-\frac{1}{2 \sqrt{2} } 
\sum_{J=3/2, 1/2}
\frac{1}{U_{J}^{e_g}} 
\left(
\frac{2}{3}\delta_{J,3/2}
+\frac{1}{3}\delta_{J,1/2}
\right)
\sum_{\substack{d_B \in e_g \\d_A \in t_{2g}}} 
\sum_{S=1,2}
\left[
\delta_{S,2} -  
\delta_{S,1} 
\right]
\left(t_{d_B, d_A}^{ \downarrow \uparrow}\right)_{S}
\\
\sum_{S'=1,2}
\left[
\delta_{S',2}\left(\frac{3}{4} + \frac{1}{\sqrt{6}}\right) +  
\delta_{S',1}\left(\frac{1}{4} - \frac{1}{\sqrt{6}}\right) 
\right]
\left(t_{d_B, d_A}^{ \uparrow \uparrow}\right)^*_{S'} .
\end{split}
\end{equation}
It describes the diagonal terms after the projection onto the three-electron intermediate spin states with the total spin $J=3/2,1/2$, which gives rise to the factor $\frac{2}{3} \delta_{J,3/2}+ \frac{1}{3}\delta_{J,1/2}$. As the result, we obtain two $t_{2g}$ electrons in  the state $\ket{1,0}$ and the spin up $e_{g}$ electron. The $t_{2g}$ vacancy can be filled by an electron with spin up or spin down projections. After projecting these two possible states on the final state,  one obtains $M_{\rm SI,3a}^{-}$. %
Calculation of the remaining matrix elements is analogous to the one discussed above. $M_{\rm SI,3b}^{-}$ is expressed as
    \begin{equation}
\label{eq:ch5:MAA3b}
\begin{split}
M_{\rm SI,3b}^{-} = 
-\frac{1}{6}
\sum_{J=3/2, 1/2}
\frac{1}{U_{J}^{e_g}} 
\left(
\delta_{J,3/2}
-\delta_{J,1/2}
\right)
\sum_{\substack{d_B \in e_g\\ d_A \in t_{2g}}} 
\sum_{S=1,2}
\left[
\delta_{S,2} - \delta_{S,1} 
\right]
\left(t_{d_B, d_A}^{ \downarrow \uparrow}\right)_{S}
\\
\sum_{S'=1,2}
\left[
\delta_{S',2}\left(\frac{\sqrt{3}}{4} + \frac{1}{\sqrt{6}}\right) +  
\delta_{S',1}\left(-\frac{\sqrt{3}}{4} + \frac{1}{\sqrt{6}}\right) 
\right]
\left(t_{d_B, d_A}^{ \downarrow \downarrow}\right)^*_{S'} .
\end{split}
\end{equation}
For $M_{\rm SI,4}^{-} = M_{\rm SI,4a}^{-} + M_{\rm SI,4b}^{-}$,  $M_{\rm SI,4a}^{-}$ is 
\begin{equation}
\label{eq:ch5:MAA4a}
\begin{split}
M_{\rm SI,4a}^{-} = 
-
\sum_{J=3/2, 1/2}
\frac{1}{U_{J}^{e_g}} 
\left(
\frac{1}{3}\delta_{J,3/2}
+\frac{2}{3}\delta_{J,1/2}
\right)
\sum_{\substack{d_B \in e_g \\d_A \in t_{2g}}} 
\sum_{S=1,2}
\left[
\frac{1}{4}\delta_{S,2} +  
\frac{3}{4}\delta_{S,1} 
\right]
\left(t_{d_B, d_A}^{ \downarrow \uparrow}\right)_{S}
\\
\sum_{S'=1,2}
\left[
\delta_{S',2}\left(\frac{\sqrt{3}}{4} + \frac{1}{\sqrt{6}}\right) +  
\delta_{S',1}\left(-\frac{\sqrt{3}}{4} + \frac{1}{\sqrt{6}}\right) 
\right]
\left(t_{d_B, d_A}^{ \downarrow \downarrow}\right)^*_{S'} ,
\end{split}
\end{equation}
and 
$M_{\rm SI,4b}^{-}$ is 
\begin{equation}
\label{eq:ch5:MAA4b}
\begin{split}
M_{\rm SI,4b}^{-} = 
-\frac{\sqrt{2}}{3}
\sum_{J=3/2, 1/2}
\frac{1}{U_{J}^{e_g}} 
\left(
\delta_{J,3/2}
-\delta_{J,1/2}
\right)
\sum_{\substack{d_B \in e_g\\ d_A \in t_{2g}}} 
\sum_{S=1,2}
\left[
\frac{1}{4}\delta_{S,2} +  
\frac{3}{4}\delta_{S,1} 
\right]
\left(t_{d_B, d_A}^{ \downarrow \uparrow}\right)_{S}
\\
\sum_{S'=1,2}
\left[
\delta_{S',2}\left(\frac{3}{4} + \frac{1}{\sqrt{6}}\right) +  
\delta_{S',1}\left(\frac{1}{4} - \frac{1}{\sqrt{6}}\right) 
\right]
\left(t_{d_B, d_A}^{ \uparrow \uparrow}\right)^*_{S'} .
\end{split}
\end{equation}
\end{widetext}

\bibliographystyle{apsrev4-2} %
\bibliography{ref} %

\end{document}